\documentclass[aps,prl,preprint,groupedaddress]{revtex4-1}
\usepackage[usenames]{color}
\usepackage{graphics}
\usepackage{epsfig}
\usepackage{epsf}
\usepackage{latexsym}
\usepackage{amsmath}
\usepackage{amssymb}
     \newcommand{\bra}[1]{\langle#1|}\newcommand{\ket}[1]{|#1\rangle}   \parskip=5pt

\graphicspath{%
    {converted_graphics/}% inserted by PCTeX
    {C:/}% inserted by PCTeX
}
\begin{document}

\title{Stochastic epidemic-type model with enhanced connectivity:  exact solution}
\author{H. T. Williams}
\email {williamsh@wlu.edu}
\author{I. Mazilu}
\email {mazilui@wlu.edu}
\author{D. A. Mazilu}
\email{mazilud@wlu.edu}
\affiliation{Department of Physics and Engineering \\ Washington and Lee University, Lexington, VA  24450, USA}
\date{\today}
 \begin{abstract}
We present an exact analytical solution to a one-dimensional model of the Susceptible-Infected-Recovered (SIR) epidemic type, with infection rates dependent on nearest-neighbor occupations. We use a quantum mechanical approach, transforming the master equation via a quantum spin operator formulation. We calculate exactly the time-dependent density of infected, recovered and susceptible populations for random initial conditions, and compare our results with a low connectivity SIR model reported by Sch\"{u}tz \it et al.   \rm  \cite{schutz}. Our results compare well to those of previous work, validating the model as a useful tool for additional and extended studies in this important area.  Our model also provides exact solutions for the n-point correlation functions, and can be extended to more complex epidemic type models.
\end{abstract}
\maketitle

\section{Introduction}

The study of cooperative evolution of multi-agent systems is a part of many sciences, from biology and social science to physics, chemistry and engineering. The methods of statistical physics are often employed, and simple models are building blocks in this quest to understand complexity. From the point of view of non-equilibrium statistical physics, biology is an exciting area of investigation. After all, every living organism is an example of a far-from-equilibrium system, and stochastic processes are ubiquitous in biological systems.

 Epidemic-type models abound in the literature\cite{book reviews}, \cite{original SIR}, from very simple ones that capture the basic rules of the infection mechanism, to very complex models that account for spatial spread, age structure and the possibility of immunization \cite{complex models}. It is interesting to see how some of these epidemic-type models have been applied  successfully in other fields as well, such as social sciences (voter models, rumor spreading models) \cite{liggett},\cite{privman} or computer science (the spread of a virus in a computer network) \cite{computer virus}. Some models are deterministic, following a set of evolution equations with given initial conditions solved using the mean field theory approach, while others are stochastic and studied using methods such as the Langevin equation, the Fokker-Planck equation and computer simulations. Some recent numerical studies of epidemic-type models can be found in \cite{numerical studies}.

Despite numerous studies and approximation schemes, exact solutions for epidemic models are rare. A study that sparked our interest was published in 2008 by Sch\"{u}tz \it et al. \rm \cite{schutz}. This presents an exact solution for a stochastic one-dimensional SIR (susceptible/infected/recovered) epidemic model. The method used is a quantum mechanical formulation of the master equation in terms of second quantized operators. The authors define cluster functions  that describe the behavior of susceptibles adjacent to infected individuals at the cluster boundaries. They derive and solve exactily a set of coupled evolution equations for these functions. Their exact solution shows the significant difference between low connectivity and high connectivity SIR models, and the role of fluctuations.  Fluctuations are built into the exact solutions, but are missing in the mean field approach.

Working towards an exact solution for an SIR model with higher connectivity, we study a variation of a one-dimensional SIR model in which we define different rates of infection depending on the number of infected neighbors. A susceptible individual with two infected neighbors will have a different probability of being infected than a susceptible neighboring only one infected person. As in the traditional SIR model, we assume the possibility of recovery. (To differentiate it from other one-dimensional SIR models, we refer to our model as the \it dual neighbor model.\rm) This model also can be cast as a non-conservative voter model, with the three classes of individuals defined as S - undecided, I - biased, R - decided.

 We here present an exact solution of the dual-neighbor SIR model. We employ a quantum mechanical approach to the problem, using the cluster function method used by Sch\"{u}tz \cite{schutz}. The steady-state solution depends on initial populations of susceptible and infected individuals. It has fluctuations built-in, and has a stationary state different from the low connectivity SIR model. Although the overall trend of the solution is similar to that of the low connectivity model, there are significant differences as well.

In Section 2, we define our model and its quantum mechanical representation. Next, we present the cluster function method and derive the evolution equations for the cluster functions and particle densities that fully resolve the model (Section 3). We conclude with an analysis of our solutions, summarize of our work, and suggest some interesting open questions (Section 4).

\section{Dual Neighbor Model and its quantum mechanical representation} 

The traditional SIR model consists of a fixed number of individuals N split into three classes: susceptible, infected, and recovered.  In a high connectivity (mean field) model, each "node" of the network is represented by an individual in one of the three classes, in contact with  every other node. A susceptible becomes infected with rate $\beta$ when in contact with an infected; an infected individual recovers spontaneously with rate $\alpha$; recovered individuals cannot change.  At a particular time the average number of individuals in each class is represented by $\bar{S}$, $\bar{I}$, and $\bar{R}$, with $\bar{S} + \bar{I} + \bar{R} = N$.  

The time evolution of these classes is governed by a set of coupled differential equations:
\begin{eqnarray*}
\frac{d\bar{S}}{dt}&=&-\beta \bar{S}\bar{I} \label{trad1} \\
\frac{d \bar{I}}{dt}&=&\beta \bar{S}\bar{I} -\alpha \bar{I} \label{trad2} \\
\frac{d\bar{R}}{dt}&=&+\alpha \bar{I} \label{trad3}
\end{eqnarray*}
The first equation describes reduction of $\bar{S}$ via infection; the second shows $\bar{I}$ increasing via infection of susceptibles, and decreasing by spontaneous recovery; and the third shows the increase of $\bar{R}$ due to spontaneous recovery.  This system of coupled nonlinear differential equations can be solved numerically, yielding the time dependence of each class of individuals, but without any information regarding correlations between individual nodes.  This model exhibits smooth time dependence of the class populations, without any statistical fluctuations.\\

In contrast, we propose a stochastic one-dimensional model in which each node is in contact with only two other nodes (as if arrayed along a line), and the rate of infection depends on the number of infected neighbors. Representing linear sequences of neighboring individuals via strings of symbols (e.g. $SIS$ representing a susceptible to the left of an infected to the left of another susceptible), the dynamics of this model is defined as:
\begin{eqnarray*}
ISI &\rightarrow& III\mbox{ with rate $\beta$}\\
ISS&\rightarrow& IIS  \mbox{ with rate $\lambda$}\\
SSI&\rightarrow& SII  \mbox{ with rate $\lambda$}\\
I&\rightarrow& R \mbox{ with rate $\alpha$}
\end{eqnarray*}
The first process describes the mechanism of infection of a susceptible neighboring two infecteds; the next two processes describe infection with a different rate when the susceptible has only one infected neighbor; and the last process describes spontaneous recovery with yet another rate.  We can pick the values of $\beta$, $\lambda$ and $\alpha$ based on the application of the model. To model actual epidemics, for example, we would assign a higher rate of infection when a susceptible is in contact with two infecteds compared to just one, thus $ \beta> \lambda$. Note that our model does not allow a succeptible individual with a recovered on one side and an infected on the other ($ISR$ or $RSI$) to become infected.  As a model of disease transmission this represents an immunity provided by a recovered neighbor; in a voter model, it suggests that an undecided voter remains so if flanked by one biased voter and one who is decided.  Practically, it says that the rate of infection in $ISR$ and $RSI$ configurations are negligible compared to other infection rates $\alpha$ and $\beta$.  Making such an assumption produces a Hamiltonian amenable to exact solution of the rate equations.  As we shall exhibit, features of the results of this model validate it as a reasonable first approximation to real-world situations.

The related mean field (deterministic)  model is governed by the following differential equations for the average populations:
\begin{eqnarray*}
\frac{d\bar{S}}{dt}&=&-\beta \bar{I} ^{2}\bar{S}-\lambda \bar{I} \bar{S}^{2} \\
\frac{d\bar{I}}{dt}&=&\beta \bar{I}^{2}\bar{S}+\lambda \bar{I} \bar{S}^{2}-\alpha \bar{I} \\
\frac{d\bar{R}}{dt}&=&+\alpha \bar{I}
\end{eqnarray*}
Compared with the traditional SIR model, the dynamics of this system is governed by three-point interactions between the S and I type individuals. This system also can be solved numerically for the time dependence of  $\bar{S}$, $\bar{I}$ and $\bar{R}$.\\

Proceeding beyond the mean field approximation, the time evolution of our model is best described by the master equation, expressing conservation of probability within a continuous-time dynamics. We let $C$ represent a configuration, giving the state ($S$, $I$, $R$) of each of the $N$ individuals.  Proximity is defined by labeling each individual with an index $i=1,2,\ldots,N$ and assuming individual $N$ is adjacent to individual $1$ (periodic boundary conditions.) The master equation expresses the rate of change of the probability $P(C ,t)$ of finding the system in configuration $C$ at time $t$ as the rate of transfer of probability into $C$ from other configurations less the rate at which $C$ passes probability into others \cite{glauber}:    
\begin{equation} 
\frac{dP(C ,t)}{dt}=\sum_{C' \neq C}\left\{ r\left[ C' \rightarrow C \right] P(C',t)
 -r\left[ C \rightarrow C'\right] P(C ,t)\right\}   \label{master}
\end{equation}
The transition rate $r\left[ C \rightarrow C' \right]$ is the probability per unit time that configuration $C$ changes into a different configuration $C'$.

Utilizing Dirac notation, we represent a configuration as $\ket{C}$, and use standard methods \cite{schutz1} to build a vector representation for a general state as a probabilistic superposition of all configurations of a system:
\begin{equation} 
\ket{P(t)}=\sum_{C}P(C,t)\ket{C} 
\end{equation}
where $P(C,t)$ is the probability that the system will be found in configuration $C$ at time $t$.
This allows the master equation to be re-written as
\begin{equation} 
\frac{d}{dt} \ket{P(t)} = -H |P(t)\rangle \label{me1}
\end{equation}
where the pseudo-Hamiltonian $H$ has matrix elements: 
\begin{eqnarray}
  \bra{C'}H\ket{C}&=& -r(C \rightarrow C'), \;C'\neq C \nonumber \\
  \bra{C}H\ket{C} &=& \sum_{C' \neq C}r(C \rightarrow C') .
\end{eqnarray}
A formal solution to Eq. \ref{me1} can be written as $\ket{P(t)} = e^{-Ht} \ket{P(0)}$. 
In this formalism, the expectation value (at time t) of a physical quantity that has value $M(C)$ for configuration $C$  is
\begin{equation}
<M>\;=\sum_{C's}P(C,t)M(C)=\;<s|\hat{M}|P(t)> ,
\end{equation}
where $\hat{M} = \sum_C M(C) |C><C|$ is the operator corresponding to the observable $M$ and the state $|s>$ is the sum (with weight 1) of all configurations.
Time dependence of $<M>$ obeys  
\begin{equation}
   \frac{\partial <M>}{\partial t}=\; <s|[\hat{M},H]|P(t)>
\end{equation}
where  $[\hat{M},H]$ is the commutator of the $M$ operator with the Hamiltonian.

The operators for our model can be written in terms of creation and annihilation operators, following the rules for constructing a quantum Hamiltonian as presented in \cite{schutz}. We define the $a_i^{\dagger}$, $a_i$ to be the creation and annihilation operators for a type $S$ particle at site $i$, and $b_i^{\dagger}$, $b_i$ to be the creation and annihilation operators for a type $I$ particle at site $i$.  A site with neither an infected nor a susceptible is assumed to contain a recovered.  Operators at different sites commute, and operators of the same species at the same site anticommute.
The number operator $A_i \equiv a^{\dagger}_{i}a_{i}$ ($B_i \equiv b^{\dagger}_{i}b_{i}$) can only take values 0 and 1, and its expectation value equals  the density of type $A$ ($B$) particles. In this notation, 
\begin{eqnarray}
-H= &\beta&  \sum_{i}[b^{\dagger}_{i}a_{i}-A_{i}(1-B_{i})]B_{i-1}B_{i+1}+\lambda \sum_{i}[b^{\dagger}_{i}a_{i}-A_{i}(1-B_{i})]A_{i-1}B_{i+1} + \nonumber \\
 &\lambda&  \sum_{i}[b^{\dagger}_{i}a_{i}-A_{i}(1-B_{i})]B_{i-1}A_{i+1}+\alpha \sum_{i}[b_{i}-B_{i}] .
\end{eqnarray}

We can rescale the time variable in such a way that the rate $\lambda$ (assumed non-zero) becomes one.  With that assumption, and defining $\gamma \equiv \frac{\beta}{\lambda}$ and $\delta \equiv \frac{\alpha}{\lambda}$, the Hamiltonian obeys  
\begin{equation}
-H=\sum_{i}[b^{\dagger}_{i}a_{i}-A_{i}(1-B_{i})][\gamma B_{i-1}B_{i+1}+A_{i-1}B_{i+1}+B_{i-1}A_{i+1}]+\delta \sum_{i}[b_{i}-B_{i}] . \label{ham}
\end{equation}

\section{Exact solution: derivation and analysis}

We seek to find the time dependence of the average particle density of each kind of individual.  To this end, we introduce $n$-point cluster functions following the method of \cite{schutz}, \cite{peschel}:
\begin{eqnarray}
K_{r}(n) \equiv \;<A_{r}A_{r+1}..A_{r+n-1}B_{r+n}> , \\
G_{r}(n) \equiv \;<B_{r-1}A_{r}..A_{r+n-1}B_{r+n}> .
\end{eqnarray}
The Hamiltonian function (Eq. \ref{ham}) is invariant relative to translation along the string of sites, so if we assume a translationally invariant initial state, all future configurations also have this property.  Translation invariance makes these clusters independent of their starting point $r$. Because a susceptible cannot change into an infected unless they are neighbors, clusters change only at their edges. 

The equations for the cluster functions are derived by calculating the respective commutators of the cluster operator products with the pseudo-Hamiltonian $H$.\\
For $n=1$, this gives 
\begin{eqnarray}
\frac{dK(1)}{dt} &=& <s|[A_{r}B_{r+1},H]|P(t)> \; = -\delta K(1)- \gamma G(1) \label{Kn=1} , \\
\frac{dG(1)}{dt} &=& <s|[B_{r-1} A_{r}B_{r+1},H]|P(t)> \; = -(\gamma+2\delta)G(1)+ 2G(2) , \label{Gn=1}
\end{eqnarray}
and for $n>1$ 
\begin{eqnarray}
\frac{dK(n)}{dt} &=& <s|[ A_{r}\ldots A_{r+n-1}B_{r+n},H]|P(t)>\; = K(n+1)-(1+\delta)K(n)- G(n) \label{Kn>1} , \\
\frac{dG(n)}{dt} &=& <s|[ B_{r-1}A_{r}\ldots A_{r+n-1}B_{r+n},H]|P(t)> \; = 2 G(n+1) - 2(1+\delta)G(n) . \label{Gn>1}
\end{eqnarray}

The equations of motion for the particle density of susceptibles and infecteds are found in a similar fashion:
\begin{eqnarray}
\frac{d<A_{r}>}{dt}&=& <s|[A_r,H]|P(t)> \; = -2K(2)-\gamma G(1) \label{dAdt} , \\
\frac{d<B_{r}>}{dt}&=& <s|[B_r,H]|P(t)> \; = 2K(2)+\gamma G(1) -\delta<B_{r}> . \label{dBdt}
\end{eqnarray}
In order to solve for the particle density of the susceptibles  and infecteds, we thus need first
to find the cluster functions.

\subsection{Solution for cluster functions}

Introducing an operator $\hat{s}$ defined by the property $\hat{s} G_n = G_{n+1}$ we can rewrite Eq.\ref{Gn>1} (for $n>1$) as
\[ \frac{dG(n)}{dt} = - 2( 1+\delta - \hat{s} ) G(n) , \]
which has a formal solution
\[ G(n) = \exp\left( - 2( 1+\delta -  \hat{s} ) t \right) G(n)_{t=0} . \]
Representing the initial density of susceptibles and infecteds as $\eta_S$ and $\eta_I$, we describe an initial state such that $G(n)_{t=0} = \eta_S^{n} \eta_I^{2}$.  From this, 
\begin{equation} G(n) = \exp\left( - 2( 1+\delta -  \hat{s} ) t \right) \eta_S^{n} \eta_I^{2} = \exp\left( - 2(1+\delta -\eta_S)  t \right) \eta_S^{n}\eta_I^{2} . \end{equation}

Eq. \ref{Gn=1} for G(1)can be rewritten as
\[ \frac{dG(1)}{dt} + (\gamma+2\delta) G(1) = 2\exp\left( - 2(\delta +  \eta_I)  t \right) \eta_S^2\eta_I^{2} , \]
and can be integrated by standard means to yield 
\begin{equation}
G(1)=\frac { \eta_S \eta_I^2 \left( 2 \eta_S e^{ (\gamma -2 n_I)t }-2+\gamma \right) 
e^{- ( \gamma+2 \delta ) t}} {\gamma-2n_I} \label{G1oft} .
\end{equation}

 Using a similar method, we find the solution for $K(n)$ to be 
\begin{equation}
K(n)=\eta_S^{n}\eta_I e^{-\frac{t}{\tau}}[1-\eta_{I}\tau(1-e^{-\frac{t}{\tau}})] , \label{Knoft}
\end{equation}
where in order to express the solution more compactly, we have defined a relaxation time $\tau \equiv \frac{1}{\delta+1-\eta_{S}}$.

\subsection{Particle densities}
The pieces are in place now to find the time dependence of the average number of susceptibles and infecteds. We assume that at $t=0$ there are no recovered individuals, therefore $\eta_S+\eta_I=1$.  Combining Eq.'s \ref{dAdt}, \ref{G1oft} and \ref{Knoft} we find
\begin{equation}
   \frac{d<A>}{dt} = -2\eta_S^2\eta_I e^{-\frac{t}{\tau}}[1-\eta_{I}\tau(1-e^{-\frac{t}{\tau}})] - 
   \gamma \frac{n_Sn_I^2}{\gamma-2n_I} [ 2 n_S e^{-\frac{2t}{\tau}} +(\gamma-2) e^{-\frac{t}{\tau'}} ]
\end{equation}
where we have defined a second relaxation time $\tau' \equiv 1/(\gamma+2\delta)$.
Integration yields
\begin{equation}
   <A> \;= \eta_S^{*} + 2 \eta_S^2\eta_I \delta\tau^2e^{-\frac{t}{\tau}} + \eta_S^2\eta_I^2 \tau(\tau +\frac{\gamma }{\gamma-2\eta_I} ) e^{-\frac{2t}{\tau}} + \eta_S \eta_I^2 \frac{\gamma(\gamma-2)}{\gamma-2\eta_I}\tau' e^{-\frac{t}{\tau'}} 
\end{equation}
where the integration constant $\eta_S^{*}$ turns out to be the stationary value of the density of susceptibles for large times.  It is evaluated by enforcing $<A>_{t=0} = \eta_S$:
\begin{equation}
 \eta_S^{*}= \eta_S \left[ 1- \eta_I \left( \eta_S \tau^2(2 \delta + \eta_I ) + \eta_I \frac{\gamma}{\gamma-2\eta_I}(\tau \eta_S+(\gamma-2)\tau'\eta_I )     \right) \right] .
\end{equation}
This is non-zero because a susceptible can be "trapped" with a recovered on either side, and be no longer susceptible to infection. 
For the density of the infecteds, we combine Equations \ref{dBdt}, \ref{G1oft} and \ref{Knoft} to get:
\begin{equation}
\frac{d<B>}{dt}+\delta <B> \; =  2 \eta_S^2\eta_I \delta\tau e^{-\frac{t}{\tau}} + 2\eta_S^2\eta_I^2 (\tau +\frac{\gamma }{\gamma-2\eta_I} ) e^{-\frac{2t}{\tau}} + \eta_S \eta_I^2 \frac{\gamma(\gamma-2)}{\gamma-2\eta_I} e^{-\frac{t}{\tau'}}  ,
\end{equation}
leading to 
\begin{equation}
  <B> \; = C e^{-\delta t} - 2\eta_S^2 \delta \tau e^{-\frac{t}{\tau}} - \frac{2\eta_S^2\eta_I^2}{\delta+2\eta_I}(\tau + \frac{\gamma}{\gamma-2 \eta_I})e^{-\frac{2t}{\tau}}  - \eta_S\eta_I^2  \frac{\gamma(\gamma-2)}{(\gamma-2\eta_I)(\gamma+\delta)}e^{-\frac{t}{\tau'}} \label{Bvst} \end{equation}
where $C$ is an integration constant to be evaluated using $<B>_{t=0}=\eta_I$:
\begin{equation} C = \eta_I + 2\eta_S^2 \delta \tau  + \frac{2\eta_S^2\eta_I^2}{\delta+2\eta_I}(\tau + \frac{\gamma}{\gamma-2 \eta_I})  + \eta_S\eta_I^2  \frac{\gamma(\gamma-2)}{(\gamma-2\eta_I)(\gamma+\delta)} . \end{equation}
Note that the steady-state value for $<B>$ is zero, due to the process of spontaneous transformation of an infected into a recovered, regardless of its neighbors.

The time dependence of the average number of recovereds can be easily found from 
\[ <R>\;=1 - <A> - <B>. \]

\subsection{Analysis of solutions}
In an actual epidemic, major concerns are the time evolution  of the number of infected individuals, and their maximum number. Basic questions include: at what time is the peak of the infection reached; what are the factors that control the maximum number of infecteds; and how long does it take for a population to fully recover? With these questions in mind, we examine the exact solution for the infected population, Eq. \ref{Bvst}. The solution depends on the initial fraction of infecteds, $\eta_I$, the parameter $\gamma$ representing the relative rate of infection with two infected neighbors vs. that for one infected neighbor, and $\delta$ representing the relative rate of spontaneous recovery vs. the rate of infection with a single infected neighbor. \

We point out two values of $\gamma$ that deserve special attention,  $\gamma=1$ and $\gamma=2$. Physically, $\gamma=1$ means that the the probability of infection is the same regardless of the number of infected neighbors.  This most closely matches the previous SIR model \cite{schutz}. The case $\gamma=2$ corresponds to an infection rate proportional to the number of infected neighbors, perhaps most closely aligned with the case of a more realistic three-dimensional medical infection model.

Two typical time behaviors of the density of infecteds are shown in Fig. 1. Figure 1(a) shows the density rising from an initially small value to a maximum, and thereafter relaxing exponentially via spontaneous recovery until all infected individuals have disappeared.  If the initial infection rate is high and/or the spontaneous recovery rate is high compared to infection rates, however, the model can exhibit time behavior that shows no peak, as in Fig. 1(b). 

Figures 2(a) and 2(b) show how infection peak height and peak time vary with the initial state infection density.  The peak height shows an increase as $\eta_I$ increases, leveling off as it must as one approaches an initial state with the majority of the population already infected.  The smaller the initially infected population, the longer delay until the maximum number of infecteds is experienced.  For $\eta_I$ values close to one, the peak will be early, if it appears at all.  Because we have rescaled time in such a way that the infection rate of a susceptible with a single infected neighbor is $1$, the proportional delay of the peak as a function of this single neighbor infection rate is hidden in these graphs.

Figure 3 shows how the spontaneous recovery rate $\delta$ controls whether a peak in $<B>$ appears or not.  As expected, for a high spontaneous recovery rate the infected population begins to decay immediately once recovery is "turned on," as might be the case at the beginning of an immunization effort in the population.  As the recovery rate drops, the figure shows the appearance of a peak that grows in size and in delay time.
 
Figure 4 shows the comparison of our solution (continuous curve) with results of the low connectivity model obtained by Sch\"{u}tz \it et al. \rm  (dashed curve) and the mean field approximation (dotted curve)  for two representative sets of parameters. There are noticeable differences among the three solutions when $\gamma$ becomes large in comparison to $\delta$ (Fig. 4(a)). These differences disappear for $\gamma$ values on the order of or smaller than $\delta$, as shown in Fig. 4(b).  We notice that the peak of the infection in the mean field approximation is higher and happens at a later time than the one predicted by our model.

The various behaviors exhibited in the figures well represent effects seen in actual situations of disease spread \cite{mollison}.  This serves to commend the model as a reasonable approach to modeling such behavior, and the exact analytical solutions enable precise predictions about features of the solutions.  

\section{Conclusions}
In this paper we analyzed a modified one dimensional stochastic SIR model, the dual neighbor model, with infection rates dependent on the nearest-neighbor occupation.  Using a quantum mechanical approach and the cluster function method introduced in Sch\"{u}tz \it et al.\ \rm \cite {schutz}, we found an exact solution for the mean densities of susceptibles, infecteds and recovereds as a function of time and initial conditions. The quantum mechanical approach is a powerful analytical tool, because it leads to exact solutions not only for the mean number of infecteds, susceptibles and recovereds, but also for n-point correlation functions.

We analyzed our solution for the density of infecteds in various parametric regimes, and also compared our solution with the low connectivity model reported in \cite {schutz}.  The qualitative behaviors exhibited by this model are straightforward and not unexpected.  On the other hand, the details that this solution presents could lead to proposals for moderating the effects of an epidemic.  This suggests the utility of further investigation of similar models, perhaps in two dimensions.  

Although this model was introduced as an epidemic model, it can also be extended for other areas of study as well, such as voter problems, computer virus dynamics, or surface deposition. It can also be generalized to include correlated initial conditions.

We hope to further our study to find exact solutions for more complicated (and realistic) SIR models that include  time delay of infection, the possibility of immunization, and also the reappearance of the disease (R can become S). Unfortunately the cluster function method fails for this latter (SIRS) model, and other mathematical avenues must be pursued. 

\section{Acknowledgments}

I. M. wants to express special thanks to the Kavli Institute for Theoretical Physics for hospitality and financial support. This
research was supported in part by the National Science Foundation under Grant No. PHY05-51164.

\begin{figure}[tbp] % float placement: (h)ere, page (t)op, page (b)ottom, other (p)age
  \centering
  % file name: H:/Summer2010/Epidemics/Fig1.eps
  \includegraphics[bb=15 13 597 780,width=5.67in,height=7.47in,keepaspectratio]{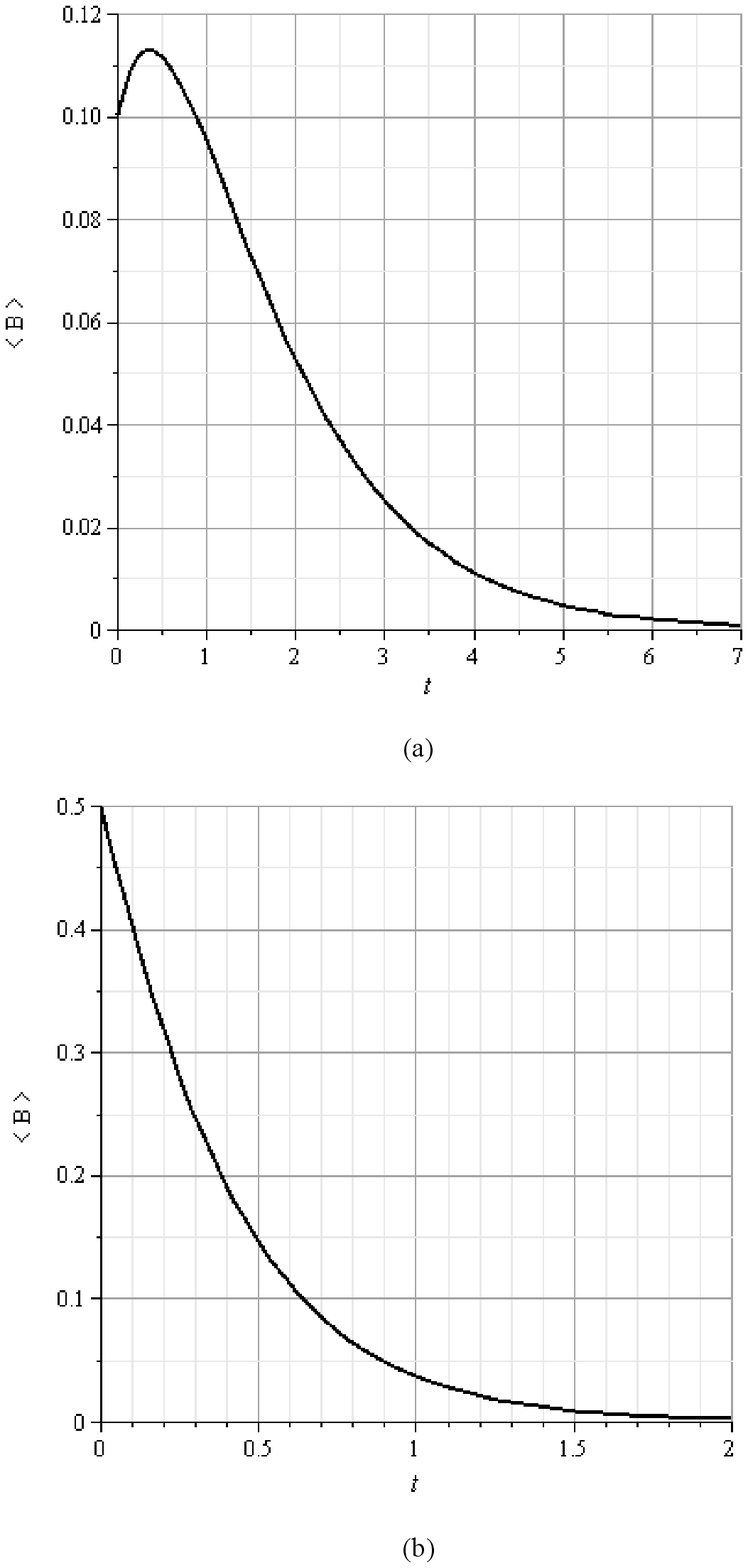}
  \caption{Typical time evolutions of the density of infected individuals $<B>$ for: (a) $\eta_{I}=0.1$, $\eta_{S}=0.9$, $\delta=1$ and $\gamma=2$; (b) $\eta_{I}=0.5$, $\eta_{S}=0.5$, $\delta=3$ and $\gamma=2$. }
 \label{fig:Fig1}
\end{figure}

\begin{figure}[tbp] % float placement: (h)ere, page (t)op, page (b)ottom, other (p)age
  \centering
  % file name: H:/Summer2010/Epidemics/Fig2.eps
  \includegraphics[bb=15 13 597 780,width=5.67in,height=7.47in,keepaspectratio]{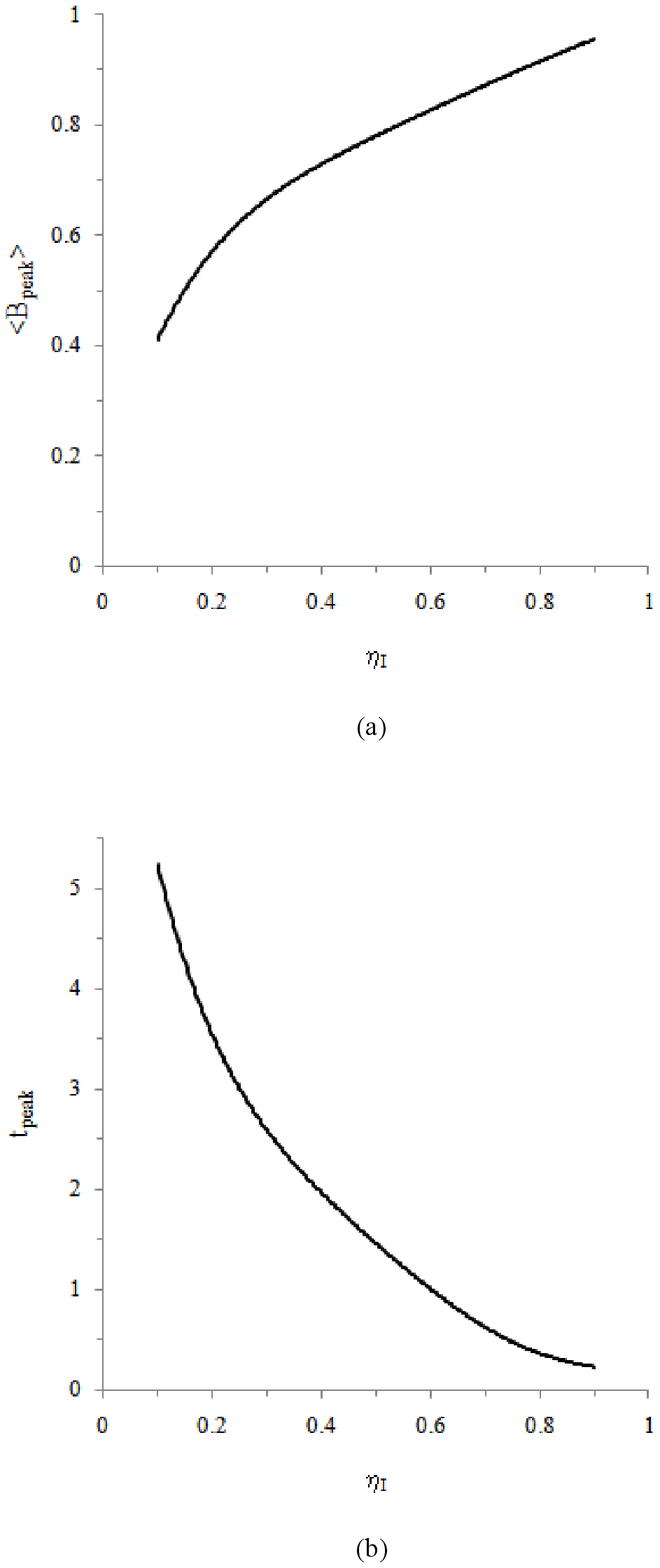}
  \caption{(a) Height of the infection peak as a function of the initial density of infected for $\gamma=2$, $\delta=0.1$; (b)  Time of the infection peak as a function of the initial density of infected for $\gamma=2$, $\delta=0.1$.}
  \label{fig:Fig2}
\end{figure}

\begin{figure}[tbp] % float placement: (h)ere, page (t)op, page (b)ottom, other (p)age
  \centering
  % file name: H:/Summer2010/Epidemics/Fig3.eps
  \includegraphics[bb=15 13 597 780,width=5.67in,height=7.47in,keepaspectratio]{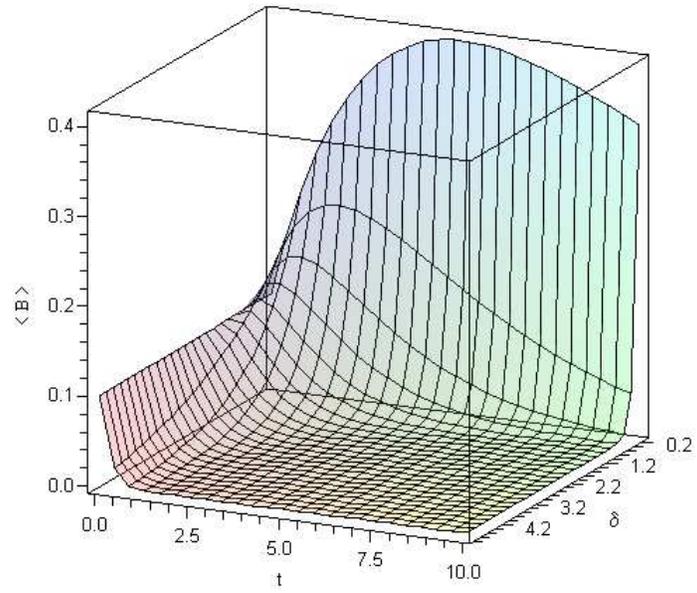}
  \caption{Density of infected individuals as a function of time and recovery rate $\delta$ for $\eta_{I}=0.1$, $\eta_{S}=0.9$ and $\gamma=2$.}
  \label{fig:Fig3}
\end{figure}

\begin{figure}[tbp] % float placement: (h)ere, page (t)op, page (b)ottom, other (p)age
  \centering
  % file name: H:/Summer2010/Epidemics/Fig4.eps
  \includegraphics[bb=15 13 597 780,width=5.67in,height=7.47in,keepaspectratio]{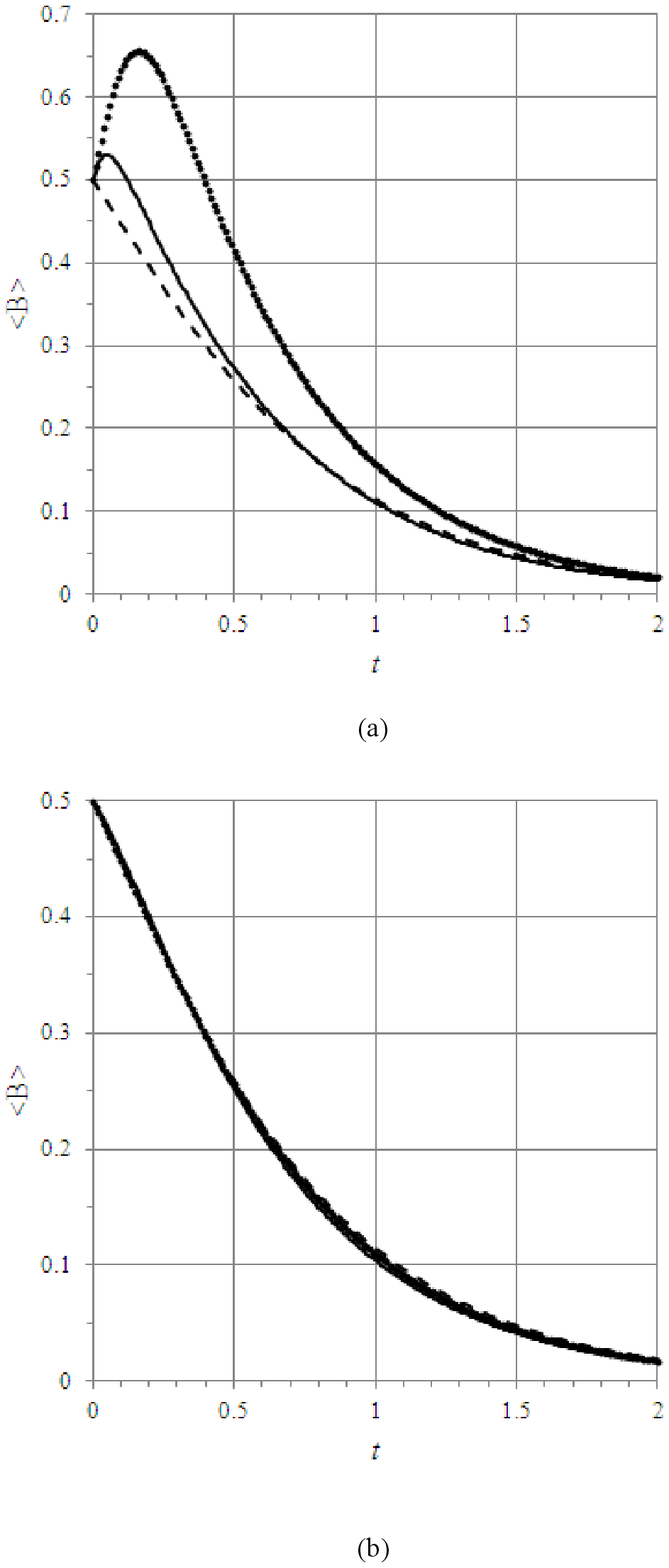}
 \caption{Comparison with the mean field model (dotted curve) and the low connectivity model (dashed curve)  \cite{schutz}: (a) For high infection rate $\gamma=20$ and $\eta_{I}=0.5$, $\eta_{S}=0.5$ and $\delta=2$ ;(b) For low infection rate $\gamma=3$ and $\eta_{I}=0.5$, $\eta_{S}=0.5$ and $\delta=2$.}
 \label{fig:Fig4}
\end{figure}

\end{document}